\documentstyle[11pt,cal97,psfig]{article}

\begin{document}

\title{Calibration of Geometric Distortion in the ACS Detectors}

\author{G.R.\ Meurer$^1$, D.\ Lindler$^2$, J.P.\ Blakeslee$^1$, 
  C.\ Cox$^3$, A.R.\   Martel$^1$, H.D.\ Tran$^1$, R.J.\ Bouwens$^4$, 
  H.C.\ Ford$^1$, M.\ Clampin$^3$, G.F.\ Hartig$^3$, M.\ 
  Sirianni$^1$, \&\ G. de Marchi$^3$}
\affil{$^1$ Department of Physics and Astronomy, The Johns Hopkins
  University, Baltimore, MD 21218\\
  $^2$ Sigma Space Corporation, Lanham, MD 20706\\
  $^3$ Space Telescope Science institute, Baltimore, MD 21218\\
  $^4$ UCO/Lick Observatory, University of California, Santa
Cruz, CA 95064}

% Notice that some of these authors have alternate affiliations, which
% are identified by the \altaffilmark after each name. The actual alternate
% affiliation information is typeset in footnotes at the bottom of the
% first page, and the text itself is specified in \altaffiltext commands.
% There is a separate \altaffiltext for each alternate affiliation
% indicated above.

% The abstract is entered in a LaTeX "environment", designated with paired
% \begin{abstract} -- \end{abstract} commands. Other environments are
% identified by the name in the curly braces.

% Poster authors ONLY may omit the abstract in order to gain a little
% more page space for the text of the poster.

\begin{abstract}
The off-axis location of the Advanced Camera for Surveys (ACS) is the
chief (but not sole) cause of strong geometric distortion in all
detectors: the Wide Field Camera (WFC), High Resolution Camera
(HRC), and Solar Blind Camera (SBC).  Dithered observations of rich star
cluster fields are used to calibrate the distortion.  We describe the
observations obtained, the algorithms used to perform the calibrations
and the accuracy achieved.  
\end{abstract}

% Keywords should be included, but they are not printed in the hardcopy.


% That's it for the front matter. On to the main body of the paper.
% We'll only put in tutorial remarks at the beginning of each section
% so you can see entire sections together.

\section{Introduction}

Images from the Hubble Space Telescope (HST) Advanced Camera for Surveys
(ACS) suffer from strong geometric distortion: the square pixels of its
detectors project to trapezoids of varying area across the field of
view.  The tilted focal surface with respect to the chief ray is the
primary source of distortion of all three ACS detectors.  In addition,
The HST Optical Telescope Assembly induces distortion as does the ACS M2
and IM2 mirrors (which are designed to remove HST's spherical
aberration).  The SBC's optics include a photo-cathode and micro-channel
plate which also induce distortion.

Here we describe our method of calibrating the geometric distortion
using dithered observations of star clusters.  The distortion solutions
we derived are given in the IDC tables delivered in Nov 2002, and
currently implemented in the STScI CALACS pipeline.  This paper is a
more up to date summary of our results than that presented at the
workshop.  An expanded description of our procedures is
given by Meurer (2002).

\section{Method}

{\bf Observations}. The ACS SMOV geometric distortion campaign consisted
of two HST observing programs: 9028 which targeted the core of 47
Tucanae (NGC 104) with the WFC and HRC, and 9027 which consisted of SBC
observations of NGC 6681.  Additional observations from programs 9011,
9018, 9019, 9024 and 9443 were used as additional sources of data, to
check the results, and to constrain the absolute pointing of the
telescope.

The CCD exposures of 47 Tucanae were designed to well detect stars on
the main sequence turn-off at $m_B = 17.5$ in each frame.  This allows
for a high density of stars with relatively short exposures.  The F475W
filter (Sloan g') was used for the CCD observations so as to minimize
the number of saturated red giant branch stars in the field. For the HRC
two 60s exposures were taken at each pointing, while for the WFC which
has a larger time overhead, only one such exposure was obtained per
pointing.  Simulated images made prior to launch, as well as archival
WFPC2 images from Gilliland et al.\ (2000) were used to check that
crowding would not be an issue.  For calibrating the distortion in the
SBC we used exposures of NGC~6681 (300s - 450s) which was chosen for the
relatively high density of UV emitters (hot horizontal branch stars).
The pointing center was dithered around each star field.  For the WFC
and HRC pointings, the dither pattern was designed so that the offsets
between all pairs of images adequately, and non-redundantly, samples all
spatial scales from about 5 pixels to 3/4 the detector size.  For the
SBC pointings, a more regular pattern of offsets is used augmented by a
series of ~5 pixel offsets.  

\begin{figure}[t]
\vspace{-3.0cm}
\centerline{\hbox{\psfig{figure=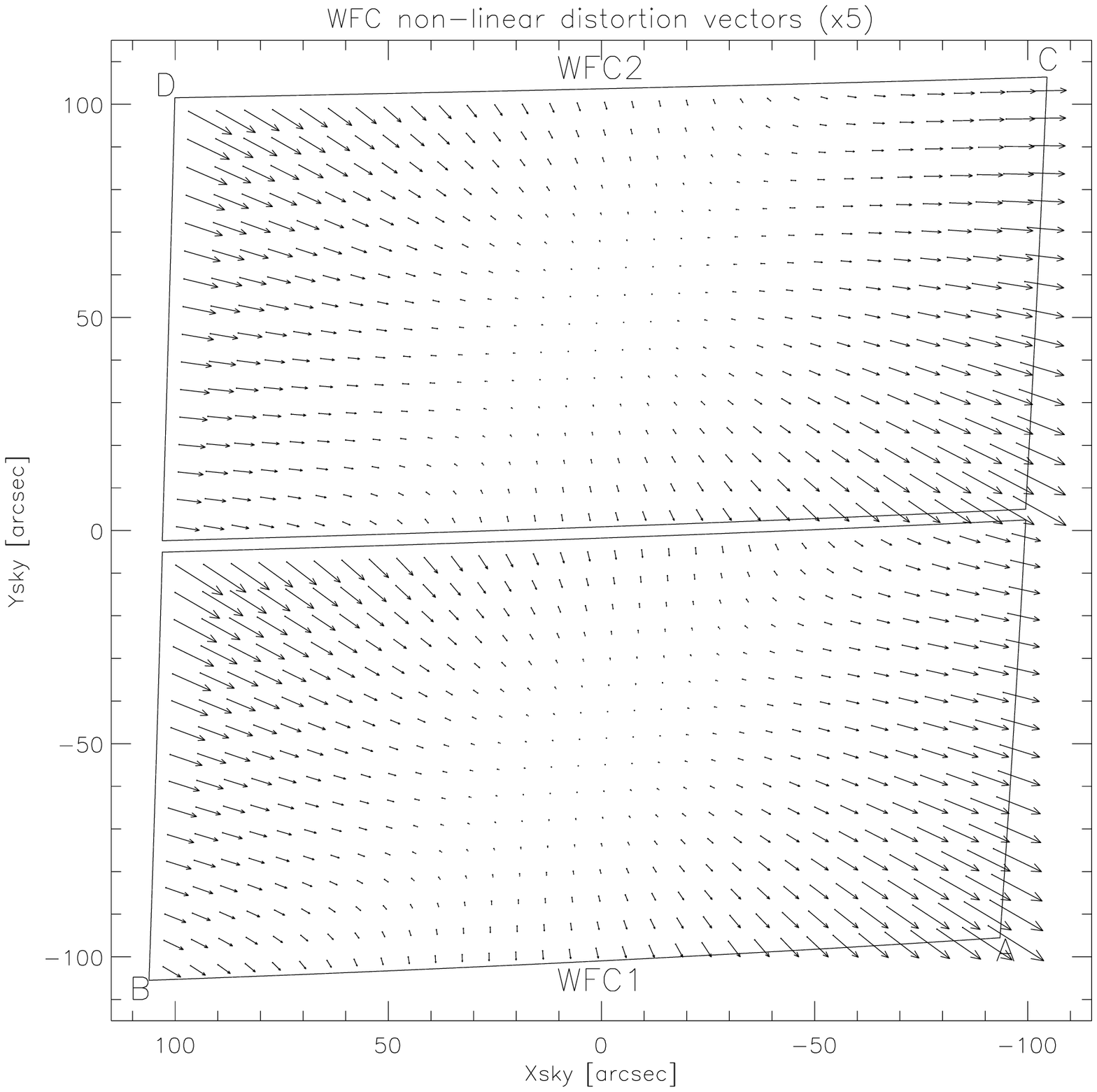,width=7.8cm}\psfig{figure=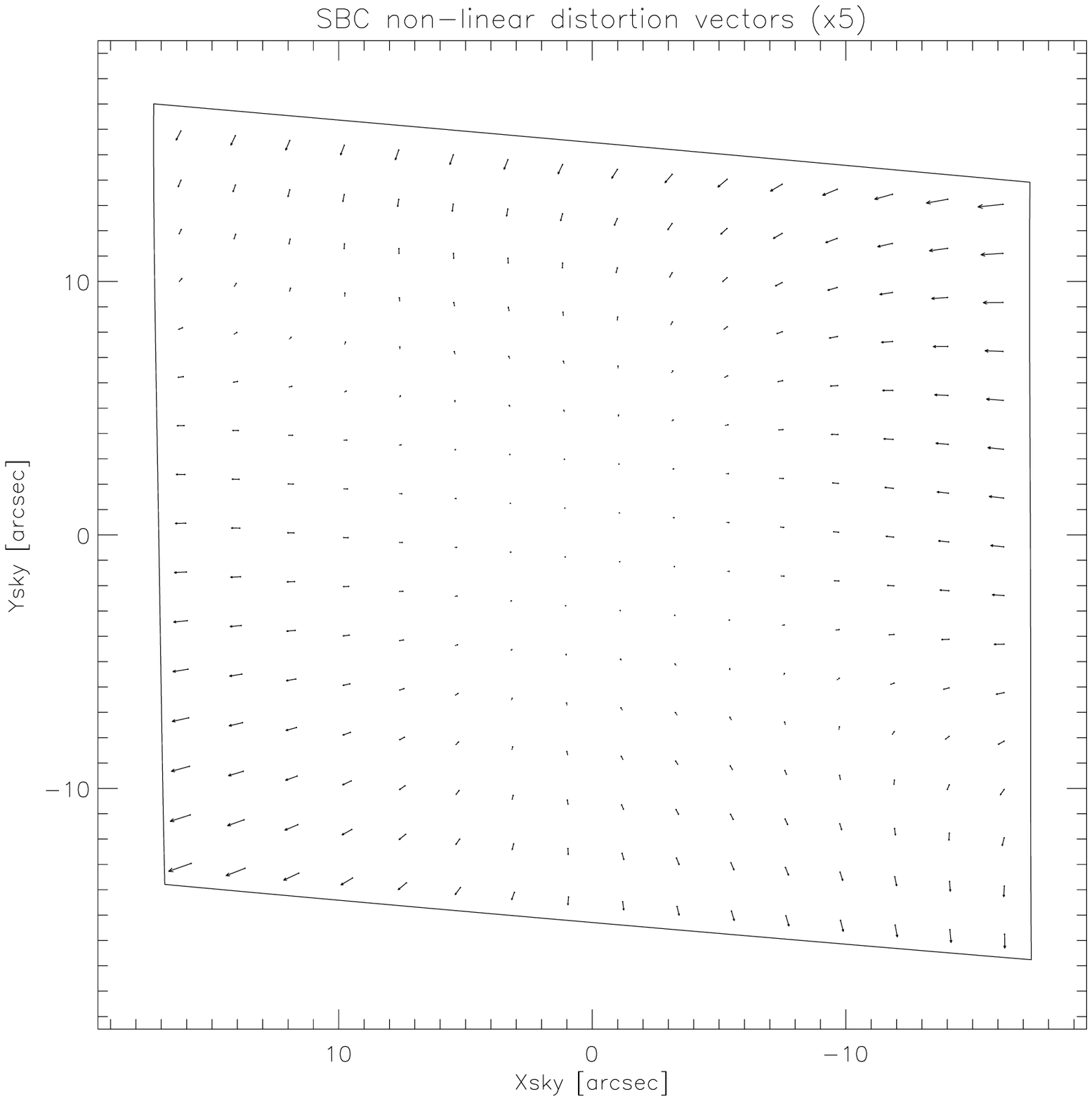,width=7.8cm}}}
\caption{Non linear component to ACS distortion for WFC and SBC
  detectors.}\label{f:nonlin}
\end{figure}

\medskip
\noindent {\bf Distortion model}. The heart of the distortion model
relates pixel position ($x,y$) to sky position using a polynomial
transformation (Hack \&\ Cox, 2000) given by: 
\begin{equation}
x_c = \sum_{m=0}^{k}\sum_{n=0}^{m} a_{m,n}(x - x_r)^n (y - y_r)^{m-n}\, , \hspace{0.5cm}
y_c = \sum_{m=0}^{k}\sum_{n=0}^{m} b_{m,n}(x - x_r)^n (y - y_r)^{m-n}
\end{equation}
Here $k$ is the order of the fit, $x_r,y_r$ is the reference pixel, taken to be the center of each
detector, or WFC chip, and $x_c,y_c$ are undistorted image coordinates.
The coefficients to the fits, $a_{m,n}$ and $b_{m,n}$, are free
parameters.  For the WFC, an offset is applied to get the two CCD chips
on the same coordinate system:
\begin{equation}
X' = x_c + \Delta{x}{\rm (chip\#)}\, , \hspace{0.5cm}
Y' = y_c + \Delta{y}{\rm (chip\#)}.
\end{equation}
$\Delta{x}{\rm (chip\#)},\Delta{y}{\rm (chip\#)}$ are 0,0 for WFC's chip
1 (as indicated by the FITS CCDCHIP keyword) and correspond to the
separation between chips 1 and 2 for chip 2.  The chip 2 offsets are
free parameters in our fit.  $X',Y'$ correspond to tangential plane
positions in arcseconds which we tie to the HST $V2, V3$ coordinate
system.  Next the positions are corrected for velocity aberration: $X =
\gamma X'$, $Y = \gamma Y'$, where
\begin{equation}
\gamma = \frac{1 + {\bf u} \cdot {\bf v} / c}{1 - (v/c)^2}.
\end{equation}
Here {\bf u} is the unit vector towards the target and {\bf v} is the
velocity vector of the telescope (heliocentric plus orbital).  Neglect
of the velocity aberration correction can result in misalignments on
order of a pixel for WFC images taken six months apart for targets near
the ecliptic.  Finally, we must transform all frames to the same
coordinate grid 
on the sky $X_{\rm sky}, Y_{\rm sky}$:
\begin{equation}
X_{\rm sky} = \cos \Delta \theta_i X - \sin \Delta \theta_i Y + \Delta X_i\, , \hspace{0.5cm} 
Y_{\rm sky} = \sin \Delta \theta_i X + \cos \Delta \theta_i Y + \Delta Y_i
\end{equation}
where the free parameters $\Delta X_i, \Delta Y_i, \Delta \theta_i$ are
the position and rotation offsets of frame $i$. 

\medskip
\noindent {\bf Calibration algorithm}. We use the positions of stars
observed multiple times in the dithered star fields to iteratively solve
for the free parameters in the distortion solution: fit coefficients
$a_{m,n}, b_{m,n}$; chip 2 offsets $\Delta x{\rm (chip\, 2)}, \Delta
y{\rm (chip\, 2)}$ (WFC only); frame offsets $\Delta X_i, \Delta Y_i,
\Delta \theta_i$; and tangential plane position $X_{\rm sky}, Y_{\rm
sky}$ of each star used in the fit.  The stars are selected by finding
local maxima above a selected threshold.  The centroid in a $7 \times 7$
box about the local maximum is compared to Gaussian fits to the $x, y$
profiles, if the two estimates of position differ by more than 0.25
pixels, the measurement is rejected as likely being effected by a cosmic
ray hit or crowding. Further details of the fit algorithm can be found
in Meurer et al.\ (2002).

\medskip
\noindent {\bf Low order terms}. Originally only SMOV images taken with
a single roll angle were used to define the distortion solutions.  The
solution using only these data is degenerate in the zeroth (absolute
pointing) and linear terms (scale, skewness).  So we used the largest
commanded offsets with a given guide star pair to set the linear terms.
However, comparison of corrected coordinates to astrometric positions
showed that residual skewness in the solution remained.  Hence, as of
November 2002, the IDC tables for WFC and SBC are based on data from
multiple roll angles.  The overall plate scale is set by the largest
commanded offset.  For the HRC, the linear scale is set by matching HRC
and WFC coordinates, since the same field was used in the SMOV
observations.  The zeroth order terms (position of the ACS apertures in
the HST $V2,V3$ frame) was determined from observations of an
astrometric field.

\section{Results}

\begin{table}[t]
\caption{Summary of fit results}\label{t:res}
\begin{center}
\scriptsize
\begin{tabular}{l c r r c r c c c}
       &      & pixel &        &           &        &        &        & \\
Camera & chip & size  & Filter & Pointings & $N$    & rms(x)\tablenotemark{1} & rms(y)\tablenotemark{1} & Notes \\
       &      & [arcsec] &     &           &        & [pixels] & [pixels] & \\
\tableline
% sigmaclip = 5.0
WFC    &  1   & 0.05  & F475W  &  25       & 142289 & 0.042  & 0.045  & \\
WFC    &  2   & 0.05  & F475W  &  25       & 103453 & 0.035  & 0.037  & \\
WFC    &  1   & 0.05  & F775W  &  10       &  31652 & 0.050  & 0.056  & 2 \\
WFC    &  2   & 0.05  & F775W  &  10       &  33834 & 0.041  & 0.048  & 2 \\
HRC    &      & 0.025 & F475W  &  20       &  77433 & 0.027  & 0.026  & \\
HRC    &      & 0.025 & F775W  &  13       &  31515 & 0.026  & 0.043  & 3 \\
HRC    &      & 0.025 & F220W  &  12       &  14715 & 0.112  & 0.108  & 3 \\
SBC    &      & 0.03  & F125LP &  34       &   1561 & 0.109  & 0.094  & \\
% No sigma clipping
% WFC    &  1   & 0.05  & F475W  &  25       & 142289 & 0.049  & 0.052  & \\
% WFC    &  2   & 0.05  & F475W  &  25       & 103453 & 0.043  & 0.044  & \\
% WFC    &  1   & 0.05  & F775W  &  10       &  31652 & 0.094  & 0.100  & 1 \\
% WFC    &  2   & 0.05  & F775W  &  10       &  33834 & 0.085  & 0.089  & 1 \\
% HRC    &      & 0.025 & F475W  &  20       &  77433 & 0.033  & 0.033  & \\
% HRC    &      & 0.025 & F775W  &  13       &  31515 & 0.037  & 0.046  & 2 \\
% HRC    &      & 0.025 & F220W  &  12       &  14715 & 0.129  & 0.125  & 2 \\
% SBC    &      & 0.03  & F125LP &  34       &   1561 & 0.127  & 0.157  & \\
\end{tabular}
\end{center}
\tablenotetext{1}{This is the rms after iteratively clipping
measurements with deviations greater than 5 times the rms. }
% without this rms clipping, the rms would be higher from factors
% ranging from 1.07 to 2.07.
\tablenotetext{2}{Coefficients held fixed to those found for WFC F475W.}
\tablenotetext{3}{Coefficients held fixed to those found for HRC F475W.}
\end{table}

The distortion in all ACS detectors is highly non-linear as illustrated
in Fig.~\ref{f:nonlin}. We find that a quartic fit ($k=4$) is adequate for
characterizing the distortion to an accuracy much better than our
requirement of 0.2 pixels over the entire field of view.
Table~\ref{t:res} summarizes the rms of the fits to the various
datasets.  

The WFC and HRC fits were all to F475W data as noted above. To check the
wavelength dependence of the distortion we used data obtained with F775W
(WFC and HRC) and F220W (HRC) from programs 9018 and 9019. We held the
coefficients fixed and only fit the offsets in order to check whether a
single distortion solution is sufficient for each detector. Table 2
shows that there is a marginal increase in the rms for the red data of
the WFC, little or no increase in the fit rms for the red HRC data, but
a significant increase in the rms using the UV data.  An examination of
the HRC F220W images reveals the most likely cause: the stellar PSF is
elongated by ~0.1". A similar elongation can also be seen in SBC PSFs.
We attribute this to aberration in the optics of either the ACS M1 or M2
mirrors or the HST OTA (Hartig, et al., 2002).  The aberration amounts to ~0.1 waves at
1600\AA, but is negligible relative to optical wavelengths, hence it is
not apparent in optical HRC images.  While it was expected that the same
distortion solution would be applicable to all filters except the
polarizers, recent work (by Tom Brown, STScI, and our team) has shown
that at least one other optical filter (F814W) induces a significant
plate scale change (factor of $\sim 4 \times 10^{-5}$).  In the long term,
the IDC tables will be selected by filter in the STScI CALACS pipeline.

\begin{figure}[t]
\vspace{-3cm}
\centerline{\hbox{\psfig{figure=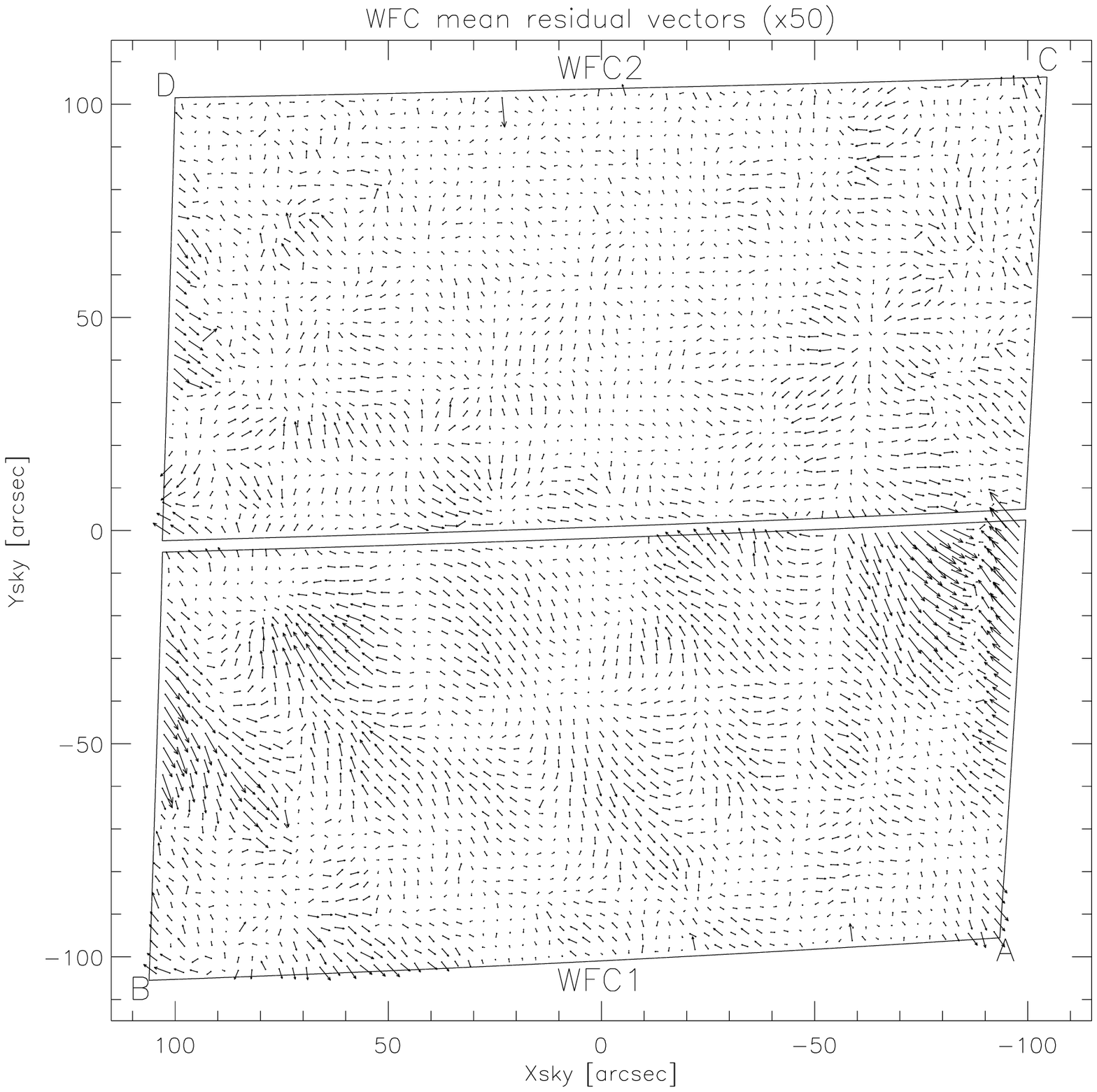,width=7.8cm}\psfig{figure=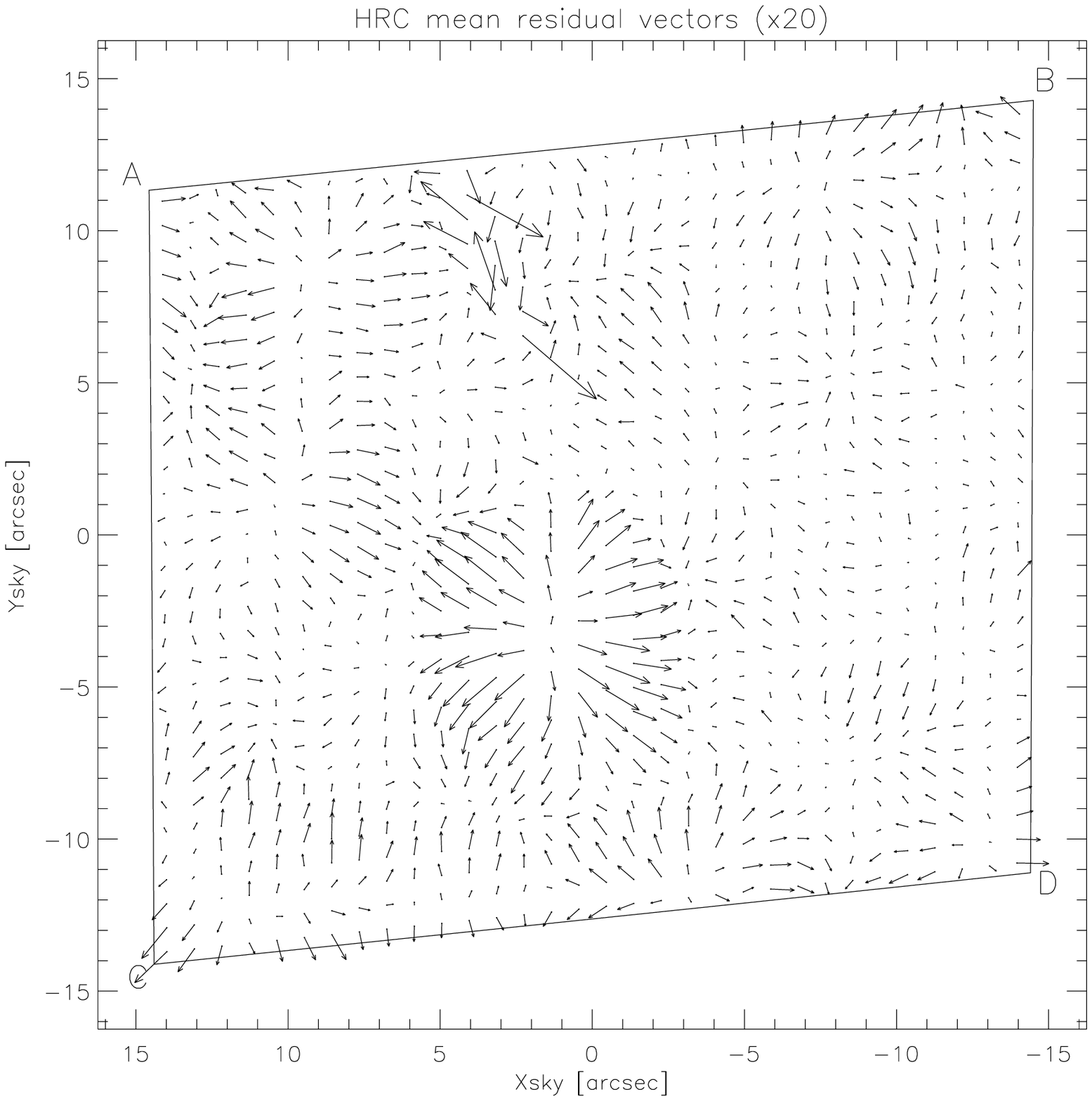,width=7.8cm}}}
\caption{Binned residuals to quartic distortion fits for the WFC and HRC
  detectors.  The large residuals in the HRC map at $X_{\rm sky} \approx
  5''\, ,\, Y_{\rm sky} \approx 10''$ coorespond to the Fastie Finger.}\label{f:resid}
\end{figure}

While a quartic solution is adequate for most purposes, binned residual
maps (Fig.~\ref{f:resid}) show that there are significant coherent
residuals in the WFC and HRC solutions.  These have amplitudes up to
$\sim 0.1$ pixels.  The small-scale geometric distortion is the subject
of the Anderson \&\ King contribution to this proceedings.  

\end{document}